\documentclass[twocolumn,showpacs,preprintnumbers,amsmath,amssymb,prb]{revtex4}
\usepackage{dcolumn}
\usepackage{bm}
\usepackage{amsmath}
\usepackage{amssymb}
\usepackage{graphicx}
\begin{document}
\title{Uniform curves for van der Waals interaction between single-wall carbon nanotubes}
\author{Evgeny G. Pogorelov}
\author{Alexander I. Zhbanov}
\email{azhbanov@gate.sinica.edu.tw}
\author{Yia-Chung Chang}
\affiliation{Research Center for Applied Sciences, Academia Sinica, 128,
  Section 2, Academia Road Nankang, Taipei 115, Taiwan.}
\date{\today}

\begin{abstract}
We report very simple and accurate algebraic expressions for the van der Waals (vdW) potentials and the forces between two parallel and crossed carbon nanotubes. The Lennard-Jones potential for two carbon atoms and the method of the smeared out approximation suggested by L.A. Girifalco were used. It is found that interaction between parallel and crossed tubes are described by different uniform curves which depend only on dimensionless distance. The explicit functions for equilibrium vdW distances, well depths, and maximal attractive forces have been given. These results may be used as a guide for analysis of experimental data to investigate interaction between nanotubes of various natures.
\end{abstract}
\maketitle

The van der Waals (vdW) interaction between graphitic structures is very important for application in Nano Electro Mechanical Systems. There are a number of publications devoted to estimation of vdW potentials for graphite layers \cite{19,20}, two fullerenes \cite{21,22,23,24}, fullerene and surface \cite{25,26}, carbon nanotube (CNT) and surface \cite{9,27,28}, fullerenes inside and outside of nanotubes \cite{29,30,31,32,33}. The interactions between the inner and the outer parallel tubes such as single- (SWNTs) \cite{29,34,35,36}, double- \cite{37,38}, and multi-wall nanotubes (MWNTs) \cite{39,40,41} are also well studied. The potential between two crossed CNTs was discussed in our previous work \cite{42}.

The continuum Lenard-Jones (LJ) model suggested by L.A. Girifalco \cite{21} is usually used to evaluate potential between two graphitic structures. The LJ potential for two carbon atoms in graphene-graphene structure is
\begin{equation}
\varphi_{C-C}(r)=-\frac{A}{r^6}+\frac{B}{r^{12}}.
\end{equation}
where $r$ is a distance, $A$ and $B$ are the attractive and repulsive constants.

The potential between two SWNTs is approximated by integration of LJ potential
\begin{equation}
\varphi=\nu^2\int\varphi_{C-C}(r)\,d\Sigma_1\,d\Sigma_2.
\end{equation}
where $d\Sigma_1$ and $d\Sigma_2$ are the surface elements for each tube. In the case of CNT-CNT interaction the mean surface density of carbon atoms is  $\nu=4/\sqrt{3}a^2$, where $a=2.46$ [\AA] is the lattice constant for graphene hexagonal structure.

The LJ potential from Eq. (2) can be integrated exactly for two crossed CNTs \cite{42}. Unfortunately the analytical formula is very complicated expression in terms of elementary functions and elliptic integrals. In the case of parallel tubes the numerical integrations were applied \cite{34,36,40}.

It was found that the vdW potential $\varphi$ between ${\rm C}_{60}-{\rm C}_{60}$, ${\rm C}_{60}$-SWNT, ${\rm C}_{60}$-graphene, graphene-graphene, and parallel SWNT-SWNT or MWNT-MWNT at different distance $d$ can be described by the universal curve \cite{29,34,40}. The universal curve for two tubes means that a plot of $\overline\varphi=\varphi/|\varphi_0|$ against $\overline d=d/d_0$ gives the same curve for all radii of tubes, where $\varphi_0$ is the minimum energy and $d_0$ is the equilibrium spacing for the two interacting surfaces.

In the present work we report very simple and accurate algebraic formulas for vdW potentials and forces between parallel and crossed SWNTs. We declare that the interaction between parallel and crossed tubes are described by different uniform curves. Also we give explicit functions for equilibrium vdW distances, potential wells, and maximal attractive forces.

\begin{figure}
\includegraphics[width=120pt]{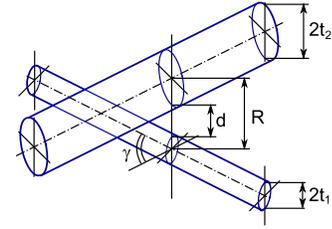}
\caption{Schematic drawing of interaction between two SWNTs.}
\end{figure}

Figure 1 illustrates the interaction between two SWNTs. In this figure $t_1$ and $t_2$ are the radii, $R$ is the distance between axes, and $d$ is the gap between surfaces of tubes. If angle $\gamma=0$ then tubes are in parallel.

Let's consider first the case of crossed tubes. In our previous work \cite{42} we obtained for two crossed SWNTs that the vdW potential is
\begin{equation}
\label{pot_tt}
\varphi=\frac{\nu^2}{\sin\gamma}
\biggl(-Ag_A+Bg_B\biggr),
\end{equation}
where $g_A$ and $g_B$ are complicated expressions in terms of elliptic functions.

These expressions can be essentially simplified when $d/t_1\ll 1$ and $d/t_2\ll 1$. Introducing $\alpha=1/t_1+1/t_2$ and using expansion in small parameter we obtain the approximating formulas
\begin{equation}
g_A=\frac{\pi^2 \sqrt{t_1 t_2}}{d^3}\biggl(\frac{1}{3}+
\frac{\alpha d}{48}\biggr),
\end{equation}
\begin{equation}
g_B=\frac{2 \pi^2 \sqrt{t_1 t_2}}{45 d^9}.
\end{equation}

This approximation allows us to explain the existence of uniform curves for crossed SWNTs. Also it allows to get simple expressions of equilibrium vdW distance $d_0$, minimum vdW potential $\varphi_0$, uniform curve $\overline\varphi=\varphi/|\varphi_0|$, and total force $F$ for interaction between two crossed SWNTs.

From Eq. (3) using Eqs. (4) and (5) we can find $\partial \varphi/\partial d=0$ and write the recurrent equation for equilibrium distance $d_0$:
\begin{equation}
d_0=\biggl(\frac{B}{A}\biggr)^{1/6}
\biggl(\frac{48}{5\alpha d_0+120}\biggr)^{1/6}.
\end{equation}

One can solve Eq. (6), passing a few steps of this recursive formula with very fast convergence.

If $\alpha d_0\ll 1$, that is $t_1$ and $t_2$ are big enough in comparison with $d_0$, then we have the first approximation for equilibrium spacing
\begin{equation}
d_0^{(1)}=\biggl(\frac{2B}{5A}\biggr)^{1/6}.
\end{equation}

Remarkable that for SWNTs of large radii, the equilibrium distance depends only from the attractive and repulsive constants $A$ and $B$. Measuring the gap $d_0$ it is possible to find the ratio between $A$ and $B$.

We find that Eq. (7) works well if $t_1,t_2>6$ [\AA] and gives small error for tubes of smallest radii. The second approximation provides high accuracy even for $t_1,t_2\approx 2$ [\AA]:

\begin{equation}
\label{pot_tt}
d_0^{(2)}(\alpha)=\biggl(\frac{48B}{A}\biggr)^{1/6}
\biggl( 5\alpha \biggl(\frac{2B}{5A}\biggr)^{1/6}
+120\biggr)^{-1/6}.
\end{equation}

If $A=15.2$ [eV$\mbox{\AA}^6$], $B=24100$ [eV$\mbox{\AA}^{12}$] as in Ref. \onlinecite{29}, and $t_1$, $t_2$ are changing from 3.40 to 20.35 [\AA], then $d_0(\alpha)$ is in the 2.914 - 2.927[\AA] range. If $t_1$ and $t_2$ tend to infinity then $d_0=d_0^{(1)}=2.931$[\AA].

Using exact analytical formula for potential from our previous work \cite{42}, we can calculate the accurate value of equilibrium distance $d_0(t_1,t_2)$. The maximal difference between exact value $d_0(t_1,t_2)$ and approximation $d_0^{(2)}(\alpha)$ does not exceed 0.1\%.

From Eqs. (3), (4), and (5) we have for the potential energy
\begin{equation}
\varphi(t_1,t_2,d)=\frac{\nu^2}{\sin \gamma}\biggl(C_1\sqrt{t_1t_2}+C_2\frac{t_1+t_2}{\sqrt{t_1t_2}}
\biggr),
\end{equation}
where $C_1=\pi^2(-A/3d^3+2B/45d^9)$ and $C_2=-A\pi^2/48d^2$ are the parameters which do not depend on the radii of tubes.

Substitution of equilibrium vdW gap $d_0$ into Eq. (3) gives us the potential well
\begin{equation}
\varphi_0(t_1,t_2)=\frac{\nu^2}{\sin \gamma}\biggl(C_1^0\sqrt{t_1t_2}+C_2^0\frac{t_1+t_2}{\sqrt{t_1t_2}}
\biggr).
\end{equation}

If we use the first approximation $d_0^{(1)}$ for equilibrium spacing (6) then $C_1^0=-A^{3/2}\sqrt{10}\pi^2/9\sqrt B$ and  $C_2^0=-A^{4/3}2^{2/3}5^{1/3}\pi^2/96B^{1/3}$.

In our previous work \cite{42} we have defined $\varphi_0(t_1,t_2)$ on the base of exact analytical formulas. After that we have described the potential $\varphi_0(t_1,t_2)$ by the same Eq. (10) and have numerically fitted $C_1^0=-0.19285$ [eV$\mbox{\AA}$] and $C_2^0=-0.05847$ [eV$\mbox{\AA}^2$]. Now we have proved this form of dependence and have analytically evaluated $C_1^0=-0.1928$ [eV$\mbox{\AA}$] and $C_2^0=-0.053$ [eV$\mbox{\AA}^2$],

\begin{table*}
\caption{Comparison of exact potential depth $|\varphi_0|$ [eV] (upper-right side) \cite{42} and approximation (lower-left side) for two SWNTs crossed at right angle}
\begin{tabular}{ccccccc}
\hline\hline\\
Tube type & (5,5) & (10,10 & (15,15) & (20,20) & (25,25) & (30,30) \\
Radius [\AA] & 3.40 & 6.79 & 10.18 & 13.57 & 16.96 & 20.35\\
\hline\\
(5,5) 3.40 & 0.761 $\backslash$ 0.785 & 1.060 & 1.277 & 1.463 & 1.627 & 1.777\\
(10,10) 6.79 & 1.039 & 1.415 $\backslash$ 1.433 & 1.726 & 1.977 & 2.200 & 2.403\\
(15,15) 10.18 & 1.256 & 1.711 & 2.069 $\backslash$ 2.080 & 2.382 & 2.651 & 2.895\\
(20,20) 13.57 & 1.442 & 1.963 & 2.373 & 2.722 $\backslash$ 2.729 & 3.037 & 3.317\\
(25,25) 16.96 & 1.606 & 2.186 & 2.643 & 3.031 & 3.376 $\backslash$ 3.380 & 3.691\\
(30,30) 20.35 & 1.755 & 2.388 & 2.887 & 3.312 & 3.688 & 4.029 $\backslash$ 4.031\\
\hline\hline
\end{tabular}
\end{table*}

Comparison between exact magnitude of well depth and approximating value for armchair SWNTs of different sizes is shown in Table I. The maximal error is 3\%.

On the base of Eqs. (7), (9), and (10) it is easy to express the uniform curve when $t_1$ and $t_2$ tend to infinity:
\begin{equation}
\overline\varphi(\overline d)=\frac{1-3\overline d^6}{2\overline d^9}.
\end{equation}

\begin{figure}
\includegraphics[width=250pt]{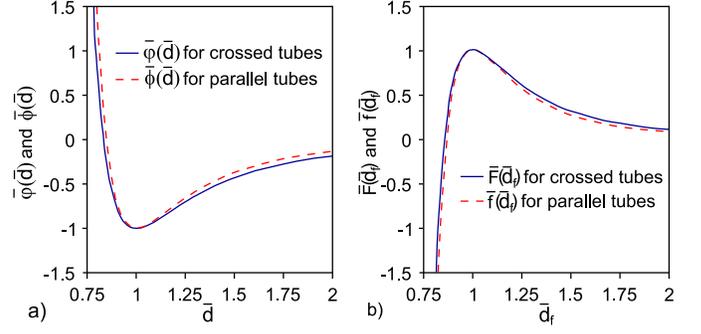}
\caption{(a) Uniform curves for potentials and (b) forces between two SWNTs.}
\end{figure}

We note that the uniform curve depends only on dimensionless distance and does not contain any material properties. The uniform curve $\overline\varphi(\overline d)$ is shown in Fig. 2(a) as a solid blue line. The plots for SWNTs of different radii fall on the uniform curve with accuracy of line thickness. The splitting between all plots does not exceed 0.013 at $\overline d=2$.

The vdW force for two crossed SWNTs can be obtained by simple differentiation $\partial \varphi/\partial d$ of vdW potential from Eq. (9):
\begin{equation}
F(t_1,t_2,d)=\frac{\nu^2}{\sin \gamma}\biggl(D_1\sqrt{t_1t_2}+D_2\frac{t_1+t_2}{\sqrt{t_1t_2}}
\biggr),
\end{equation}
where $D_1=\pi^2(A/d^4-2B/5d^{10})$ and $D_2=A\pi^2/24d^3$.

Using Eq. (12) we can find the distance $d_{\max}$ where the attractive vdW force reaches maximum. If $t_1$ and $t_2$ tend to infinity in the first approximation we have:
\begin{equation}
d_{\max}^{(1)}=\biggl(\frac{B}{A}\biggr)^{1/6}.
\end{equation}

The second approximation of higher accuracy is:
\begin{equation}
d_{\max}^{(2)}(\alpha)=\biggl(\frac{B}{A}\biggr)^{1/6}
\biggl( \frac{\alpha}{32} \biggl(\frac{B}{A}\biggr)^{1/6}
+1\biggr)^{-1/6}.
\end{equation}

If we use the first approximation (13) for $d_{\max}$ in Eq. (12) then we have the maximal attractive force:
\begin{equation}
F_{\max}(t_1,t_2)
=\frac{\nu^2}{\sin \gamma}\biggl(D_1^{\max}\sqrt{t_1t_2}+
D_2^{\max}\frac{t_1+t_2}{\sqrt{t_1t_2}}\biggr),
\end{equation}
where $D_1^{\max}=3A^{5/3}\pi^2/5B^{2/3}$ and $D_2^{\max}=A^{3/2}\pi^2/24 \sqrt B$.

If we define the uniform curve for vdW force as a plot of $\overline F= F/|F_{\max}|$ against $\overline d_f=d/d_{\max}$ then we have:
\begin{equation}
\overline F(\overline d_f)=\frac{5\overline d_f^6-2}{3\overline d_f^{10}}.
\end{equation}

The uniform curve for vdW force between two crossed SWNTs is shown in Fig. 2(b) as a solid blue line.

Let's consider the vdW interaction between two parallel SWNTs.

Using expansion in small parameter for integral (2) we get approximating formula where only the main term of the expansion is taken into account:
\begin{equation}
\phi(t_1,t_2,d)=\nu^2 \frac{\pi^2\sqrt{2t_1t_2}}{\sqrt{t_1+t_2}}\biggl(-\frac{5A}{32d^{7/2}}+\frac{2431B}{65536d^{19/2}}\biggr),
\end{equation}
here $\phi$ is expressed in units of energy per unit of length.

Analogically to previous part of present work we can calculate the equilibrium vdW distance, the potential well, and the maximal attractive force for interaction between two parallel SWNTs.

The equilibrium distance is
\begin{equation}
d_0=\frac{1}{4}\biggl(\frac{92378}{35}\biggr)^{1/6}\biggl(\frac{B}{A}\biggr)^{1/6}.
\end{equation}

If we use the lattice constant $a=2.49$ [\AA], the attractive $A=15.1636$  [eV$\mbox{\AA}^6$] and repulsive constants $B=24052$ [eV$\mbox{\AA}^{12}$] as in work \cite{34} then the equilibrium gap is $d_0=3.174$ [\AA].

The equilibrium vdW potential per unit of length is
\begin{multline}
\phi_0(t_1,t_2)=\phi(t_1,t_2,d_0)\\
=-\nu^2 \frac{120\pi^2}{877591}92378^{5/12}35^{7/12}\frac{A^{19/12}}{B^{7/12}}\sqrt{\frac{2t_1t_2}{t_1+t_2}}.
\end{multline}

Following to work \cite{34} we calculated the well depth for SWNTs of different radii from 2 to 22[\AA].

\begin{table*}
\caption{Comparison of numerically calculated potential depth $|\phi_0|$ [eV/\AA] (upper-right side) \cite{34} and our approximation (lower-left side) for two parallel SWNTs}
\begin{tabular}{ccccccc}
\hline\hline\\
Radius [\AA] & 2 & 6 & 10 & 14 & 18 & 22\\
\hline\\
2 & 50.88 $\backslash$ 48.19 & 61.82 & 67.40 & 70.53 & 72.58 & 74.00\\
6 & 62.31 & 88.12 $\backslash$ 84.74 & 95.63 & 102.27 & 106.76 & 110.01\\
10 & 65.68 & 98.53 & 113.77 $\backslash$ N/A & 119.66 & 126.26 & 131.17\\
14 & 67.31 & 104.27 & 122.88 & 134.61 $\backslash$ 131.26 & 139.62 & 145.97\\
18 & 68.26 & 107.93 & 129.00 & 142.78 & 152.64 $\backslash$ 149.47 & 157.07\\
22 & 68.89 & 110.47 & 133.40 & 148.82 & 160.09 & 168.74 $\backslash$ 165.76\\
\hline\hline
\end{tabular}
\end{table*}

Comparison between exact magnitude of well depth \cite{34} and approximating value is shown in Table II. We can conclude that the maximal difference consists 6.9\%.

The equation for uniform curve is
\begin{equation}
\overline\phi(\overline d)=\frac{7-19\overline d^6}{12\overline d^{19/2}},
\end{equation}

The uniform curve is shown in Fig 2(a) as a dotted red line. The splitting between all plots for SWNTs of different radii does not exceed 0.015 at $\overline d=2$. Remarkable that the universal curve proposed by Girifalco L.A. et al. \cite{29}
\begin{equation*}
\overline\phi_G(\overline d)=-\frac{1}{0.6}\biggl[\biggl(\frac{3.41}{3.13\overline d+0.28}\biggr)^4-0.4\biggl(\frac{3.41}{3.13\overline d+0.28}\biggr)^{10}\biggr]
\end{equation*}
is in very good agreement with our Eq. (20). If $\overline d>1$ then $\max|\overline\phi_G(\overline d)-\overline\phi(\overline d)|=0.019$ at $\overline d=1.7$.

According to Girifalco L.A et al. \cite{29} the universal curve is the plot of $\overline\phi=\phi/|\phi_0|$ against $\overline d=(R-t_1-t_2-\delta)/( R_0-t_1-t_2-\delta)$, where $R$ is the distance between centers of graphitic structures, $R_0$ is the equilibrium spacing at the minimum energy for the two interacting entities, $t_1$ and $t_2$ are the radii of tubes or fullerenes, and $\delta$ is some parameter. For interaction between two parallel SWNTs one has $\delta=0$. In other cases of interaction between graphitic structures (${\rm C}_{60}-{\rm C}_{60}$, ${\rm C}_{60}$-SWNT etc.) the fitting parameter $\delta$ is used to adjust a plot to the universal curve.
We prefer to not use any fitting parameter at all. Thus we have two different uniform curves for parallel and crossed SWNTs (see Fig. 2).

The vdW force per unit of length for two parallel SWNTs is
\begin{equation}
f(t_1,t_2,d)=\nu^2\pi^2\sqrt{\frac{2t_1t_2}{t_1+t_2}}
\biggl(\frac{35A}{64d^{9/2}}+\frac{46189B}{131072d^{21/2}}
\biggr).
\end{equation}

The distance where the attractive vdW force reaches maximum is
\begin{equation}
d_{\max}=\frac{46189^{1/6}480^{5/6}}{960}\biggl(\frac{B}{A}\biggr)^{1/6}.
\end{equation}

The maximal attractive force per unit of length is
\begin{equation}
f_{\max}(t_1,t_2)=\nu^2\frac{80\pi^2 2^{1/4} 15^{3/4}}{46189^{3/4}} \frac{A^{7/4}}{B^{3/4}}\sqrt{\frac{2t_1t_2}{t_1+t_2}}.
\end{equation}

If we define the uniform curve for vdW force between two parallel SWNTs as a plot of $\overline f= f/|f_{\max}|$ against $\overline d_f=d/d_{\max}$ then we have:
\begin{equation}
\overline f(\overline d_f)=\frac{7\overline d_f^6-3}{4\overline d_f^{21/2}}.
\end{equation}

This uniform curve is shown in Fig 2(b) as a dotted red line.

In summary, we applied Lennard-Jones potential and method of the smeared out approximation suggested by L.A. Girifalco to study interaction between two SWNTs of different diameters. Using expansion in small parameter we obtained very simple and accurate algebraic expressions of the vdW potential and the force for two parallel and crossed carbon nanotubes. It is found that interaction between parallel and crossed tubes are described by different uniform curves. The expressions of universal curves contain only dimensionless distance as a parameter and do not depend on other factors. We gave the explicit functions for equilibrium vdW distance, well depth, and maximal attractive force. We plotted uniform potential curves and uniform force curves for SWNTs. These results may be used as a guide for analysis of experimental data to investigate interaction between nanotubes of various natures.

We gratefully acknowledge support through the National Science Council of Taiwan,
Republic of China, through the project NSC 95-2112-M-001-068-MY3.
\bibliography{uniform_SWNT_new}
\end{document}